\documentclass[aps,pra,10pt,reprint,superscriptaddress]{revtex4-2}
\usepackage{amsmath,amsfonts,amssymb}
\usepackage{xcolor,graphicx}
\usepackage{braket}
\usepackage[pdftex]{hyperref}
\hypersetup{
	colorlinks = true,
	linkcolor = blue,
	anchorcolor = blue,
	citecolor = red,
	filecolor = blue,
	urlcolor = blue}
\begin{document}
\title{Lewis–Ermakov approach for the time–dependent two–level system}
\author{M. Huerta-Sandoval}
\email[e-mail:\,]{montserrat.huerta@inaoe.mx}
\affiliation{Instituto Nacional de Astrofísica Óptica y Electrónica\\ Calle Luis Enrique Erro No. 1, Santa María Tonantzintla, Puebla, 72840, Mexico}
\author{I. Ramos-Prieto}
\email[e-mail:\,]{iran@inaoep.mx}
\affiliation{Instituto Nacional de Astrofísica Óptica y Electrónica\\ Calle Luis Enrique Erro No. 1, Santa María Tonantzintla, Puebla, 72840, Mexico}
\author{H. M. Moya-Cessa}
\affiliation{Instituto Nacional de Astrofísica Óptica y Electrónica\\ Calle Luis Enrique Erro No. 1, Santa María Tonantzintla, Puebla, 72840, Mexico}
\begin{abstract}
We construct an explicit Lewis–Ermakov–type dynamical invariant for time–dependent two–level systems by exploiting their algebraic correspondence with the time–dependent harmonic oscillator through the $su(2)$ and $su(1,1)$ algebras, which share the common complexification $sl(2,\mathbb{C})$. This invariant provides a closed–form evolution operator and an exact propagator for arbitrary time–dependent coupling $a(t)$ and detuning $b(t)$. We illustrate the method on representative scenarios, including Landau–Zener transitions, non–Hermitian dissipative processes, and adiabatic rapid passage, and we obtain inverse–engineered shortcuts to adiabaticity with fully explicit control fields when the auxiliary scaling function is taken to be real.
\end{abstract}
\maketitle
\section{Introduction}
Dynamical invariants offer a robust reference framework for quantum systems governed by time–dependent Hamiltonians and play a key role in quantum control and in designing shortcuts to adiabaticity. The time–dependent harmonic oscillator serves as the standard example. For the quadratic Hamiltonian
\begin{equation}\label{H_THO}
    \hat{H}_{\texttt{THO}}(t) = \frac{a(t)}{2}\bigl(\hat{p}^2+\hat{x}^2\bigr) + \frac{b(t)}{2}\bigl(\hat{x}\hat{p}+\hat{p}\hat{x}\bigr),
\end{equation}
the Lewis–Ermakov method constructs a Hermitian invariant $\hat{I}$ satisfying
\begin{equation}
    \frac{\partial \hat{I}}{\partial t} - i[\hat{I},\hat{H}] = 0.
\end{equation}
The Schr\"odinger equation can thus be solved in terms of the time-independent eigenstates of $\hat{I}$, despite the fact that the Hamiltonian itself is explicitly time dependent~\cite{Leach_1977,Leach_1978}. The associated Ermakov–Pinney equation appears as the consistency requirement for the scaling function and underpins the analytical framework of modern inverse engineering and shortcuts to adiabaticity~\cite{Chen_2010b,Chen_2011,Torrontegui_2013,GueryOdelin_2019,Berry_2009}. Experimental realizations with trapped ions~\cite{An_2016}, superconducting circuits~\cite{Vepsalainen_2019}, and nitrogen-vacancy centers~\cite{Zhou_2017} demonstrate the practical utility of this oscillator-based approach.

In contrast, explicitly time–dependent two–level systems are typically analyzed as a distinct case. The Landau–Zener–Majorana–St\"uckelberg transition~\cite{Landau_1932, Zener_1932,Majorana_1932, Stuckelberg_1932, Shevchenko_2010} provides the standard nonadiabatic paradigm: a linear sweep of the detuning drives population transfer through an avoided level crossing. While closed-form expressions for transition probabilities are available in certain idealized regimes~\cite{Vitanov_1996}, a fully developed Lewis–Ermakov framework---featuring a dynamical invariant, a factorized time–dependent Hamiltonian, and a closed-form propagator valid for arbitrary coupling $a(t)$ and detuning $b(t)$---has not yet been formulated in the unified language used for $\hat{H}_{\texttt{THO}}(t)$. The situation becomes even more intricate in open or effectively non–Hermitian settings~\cite{Moiseyev_2011, Bender_2007}, where exceptional points and state-dependent gain modify both the spectrum and the very concept of adiabatic following.

This work fills that gap by rendering the algebraic relationship between the two systems both explicit and practically applicable. The continuous operators $\{\hat{x}^2,\hat{p}^2,\hat{x}\hat{p}+\hat{p}\hat{x}\}$ generate the non-compact algebra $su(1,1)$, whereas the discrete operators $\{\hat{\sigma}_-,\hat{\sigma}_+,\hat{\sigma}_z\}$ generate the compact algebra $su(2)$; after complexification, both give rise to the same Lie algebra $sl(2,\mathbb{C})$~\cite{Hong_2020, Lai_1996}. We show that the time–dependent two–level Hamiltonian
\begin{equation}\label{H_TLS}
    \hat{H}_{\texttt{TLS}}(t) = a(t)(\hat{\sigma}_+ + \hat{\sigma}_-) +
        b(t)\hat{\sigma}_z,
\end{equation}
can be mapped, via a pair of time–dependent transformations $\hat{G}(t)$ and $\hat{T}(t)$~\cite{Ramos_2018b,Ramos_2023,Ramos_2020,Huerta_2025}, onto the stationary form
\begin{equation}\label{H_S}
    \hat{H}_{\mathbf{S}}(t) = \frac{\theta_0}{\rho^2(t)}\bigl(\hat{\sigma}_+
        + \hat{\sigma}_-\bigr),
\end{equation}
as long as an auxiliary function $\rho(t)$ satisfies a complex Ermakov–Pinney equation whose time–dependent frequency $\theta^2(t)$ encodes $a(t)$, $b(t)$, and their time derivatives. This procedure provides an explicit dynamical invariant $\hat{I}_{\texttt{TLS}}(t)$ and a closed-form representation of the evolution operator. From a control perspective, it realizes an invariant-based shortcut protocol: by prescribing $\rho(t)$ and imposing that $\hat{I}_{\texttt{TLS}}(t)$ commute with $\hat{H}_{\texttt{TLS}}(t)$ at the initial and final times, we derive analytic control functions $a(t)$ and $b(t)$, which become fully explicit when $\rho(t)$ is chosen to be real---although, in general, our formulation permits $\rho(t)$ to depend on a complex variable.

The manuscript is organized as follows. Sec.~\ref{seccion_1} establishes the mapping between $\hat{H}_{\texttt{THO}}(t)$ and $\hat{H}_{\texttt{TLS}}(t)$, and derives both the dynamical invariant and the evolution operator. Sec.~\ref{sec:examples} applies this framework to the Landau–Zener–Majorana–St\"uckelberg model and to its dissipative generalization with an imaginary detuning $b(t)=i\alpha t$~\cite{Ivakhnenko_2023}, where the analytic predictions are compared with numerical solutions of the Schr\"odinger equation and both static and dynamic exceptional points are analyzed. The same section also examines adiabatic rapid passage in the Allen–Eberly model~\cite{Allen_1975}, illustrating how the complex Ermakov–Pinney approach enables robust population inversion and adiabatic state transfer in driven two–level systems. Sec.~\ref{sec:shortcuts} introduces inverse-engineering conditions that eliminate quantum friction and presents closed-form driving protocols based on a simple ansatz for $\rho(t)$. The appendices compile the classical and quantum derivations for the oscillator, along with the detailed transformation rules associated with the complexification of the corresponding Lie algebra. Finally, Sec.~\ref{conclusions} recaps the main findings.

\section{time–dependent two–level system}\label{seccion_1}
The quantum harmonic oscillator and the two–level system implement distinct representations of the same complexified algebra, $sl(2,\mathbb{C})$: the oscillator realizes the non-compact $su(1,1)$ through the set $\{\hat{x}^2, \hat{p}^2, \hat{x}\hat{p}+\hat{p}\hat{x}\}$, whereas the two–level system realizes the compact $su(2)$ through $\{\hat{\sigma}_-, \hat{\sigma}_+, \hat{\sigma}_z\}$. This algebraic equivalence enables one to use an analogous factorization scheme for both $\hat{H}_{\texttt{THO}}(t)$ and $\hat{H}_{\texttt{TLS}}(t)$. It is important to note that all explicit calculations are carried out within their respective real subalgebras; the complexified algebra $sl(2,\mathbb{C})$ functions only as a conceptual link and is never constructed explicitly. For the oscillator, the time dependence is extracted by means of two sequential unitary transformations, $\hat{G}(t)$ and $\hat{T}(t)$~\cite{Ramos_2018b,Ramos_2023,Ramos_2020,Huerta_2025}, under the condition that the auxiliary function $\rho(t)$ obeys the Ermakov–Pinney equation $\ddot{\rho} + \omega^2(t)\rho = \omega_0^2/\rho^3$~\cite{Ermakov_1880, Pinney_1950,Leach_1977,Leach_1978}. The Hamiltonian is then mapped to the time-independent form $\hat{\mathcal{H}}_{\mathbf{S}} = \omega_0(\hat{x}^2 + \hat{p}^2)/(2\rho^2)$, with all temporal variation captured by the scaling function $\rho(t)$. Further details are provided in Appendix~\ref{A_A}.

For the two–level system governed by Eq.~\eqref{H_TLS}, $a(t)$ characterizes the level coupling, while $b(t)$ determines the detuning. Exploiting the structural analogy with the harmonic oscillator, the time dependence is isolated using the operators $\hat{G}(t)$ and $\hat{T}(t)$. After performing the corresponding algebraic mappings, these operators take the form $\hat{G}(t) = \exp\left[i a^{-1}\left(\frac{\dot{a}}{2a} + \frac{\dot{\rho}}{\rho} + i b\right)\hat{\sigma}_-\right]$ and $\hat{T}(t) = \exp\left[\ln\left(\sqrt{\frac{a}{\theta_0}}\rho\right)\hat{\sigma}_z\right]$, with $\theta_0 = \theta(t_0)$ denoting the initial value of the frequency parameter. Because $\hat{\sigma}_-$ is both nilpotent and non–Hermitian, $\hat{G}(t)$ is inherently non-unitary. Consequently, the factorization proceeds through an intermediate, complexified state space. In this framework, the combined transformation $\hat{S}_{\texttt{TLS}}(t) = \hat{G}(t)\,\hat{T}(t)$ maps the original two–level system to a rescaled representation in which the time dependence becomes explicitly separable. It is worth stressing that, although the intermediate similarity transformation $\hat{G}(t)$ is non-unitary---precisely because of the nilpotent nature of the lowering operator $\hat{\sigma}_-$---its role is exclusively algebraic. The embedding into the complex Lie algebra $sl(2,\mathbb{C})$ supplies a mathematical setting that enables the decoupling of the differential equations. After the full time evolution is obtained and the inverse transformation $\hat{S}_{\texttt{TLS}}^{-1}(t_0)$ is applied, the physical state vector is recovered and its usual normalization is preserved.

To clarify this procedure, begin by examining how the operator $\hat{G}(t)$ incorporates the detuning term $b(t)$ to produce the Hamiltonian $\hat{H}_{G}(t)$:
\begin{equation}
    \begin{split}
        \hat{H}_{G}(t) &= \hat{G}^{-1}\,\hat{H}_{\texttt{TLS}}\,\hat{G} - i\hat{G}^{-1}\,\partial_t\hat{G},\\
        &= a\,\hat{\sigma}_+ + i\left(\frac{\dot{a}}{2a}+\frac{\dot{\rho}}{\rho}\right)\hat{\sigma}_z\\
        &+\frac{1}{a}\left(\frac{\ddot{\rho}}{\rho}+a^2 + b^2 + i\dot{b} - i\frac{\dot{a}b}{a} + \frac{\ddot{a}}{2a} - \frac{3\dot{a}^2}{4a^2}\right)\hat{\sigma}_-.
    \end{split}
\end{equation}
In this form, the $\hat{\sigma}_z$ contribution carries an imaginary prefactor determined by the driving amplitude $a(t)$ and the scaling function $\rho(t)$. The remaining time–dependent structure is collected in the coefficient of $\hat{\sigma}_-$. Next, we apply the operator $\hat{T}(t)$ to remove the $\hat{\sigma}_z$ term, obtaining the transformed Hamiltonian $\hat{H}_{T}(t)$:
\begin{equation}\label{H_T}
    \begin{split}
        \hat{H}_{T}(t)&= \hat{T}^{-1}\,\hat{H}_{G}\,\hat{T} - i\hat{T}^{-1}\,\partial_t\hat{T},\\
        &= \frac{\theta_0}{\rho^2}\,\hat{\sigma}_+\\
        &+\frac{\rho}{\theta_0}\left(\frac{\ddot{\rho}}{\rho}+a^2 + b^2 + i\dot{b} - i\frac{\dot{a}b}{a} + \frac{\ddot{a}}{2a} - \frac{3\dot{a}^2}{4a^2}\right)\hat{\sigma}_-,
    \end{split}
\end{equation}
where the $su(2)$ transformation identities summarized in Appendix~\ref{A_C} have been employed. Imposing equality between the coefficients of $\hat{\sigma}_+$ and $\hat{\sigma}_-$ to obtain a symmetric Hamiltonian requires that $\rho(t)$ obey the Ermakov–Pinney equation~\cite{Ermakov_1880, Pinney_1950}:
\begin{equation}\label{Ermakov_TLS}
    \ddot{\rho} + \theta^2(t)\rho = \frac{\theta_0^2}{\rho^3},
\end{equation}
with the effective frequency $\theta^2(t)$ given by
\begin{equation}\label{theta}
    \theta^2(t) = a^2 + b^2 + i\dot{b} - i\frac{\dot{a}b}{a} + \frac{\ddot{a}}{2a} - \frac{3\dot{a}^2}{4a^2},
\end{equation}
and $\theta_0 = \theta(t_0)$. As will be discussed later in the framework of shortcuts to adiabaticity, this constant $\theta_0$ must be adjusted or replaced by $\vartheta_0$ to fulfill the boundary conditions required for implementing the shortcut. The appearance of imaginary contributions in $\theta^2(t)$ and the opposite sign of the $b^2$ term relative to the harmonic oscillator case stem directly from the particular $su(2)$ commutation relations. Inserting Eq.~\eqref{Ermakov_TLS} into Eq.~\eqref{H_T} reproduces the scaled Hamiltonian introduced in the Introduction, Eq.~\eqref{H_S}. This time–dependent factorization allows one to solve the Schr\"odinger equation in the scaled frame and subsequently map the obtained solution back to the original laboratory frame.

Moreover, the composite transformation $\hat{S}_{\texttt{TLS}}(t)=\hat{G}(t)\,\hat{T}(t)$ does more than generate the temporal factorization that produces $\hat{H}_{\mathbf{S}}(t)$; it also specifies the dynamical invariant governing the driven two–level system. This invariant is obtained by acting with the transformation on the stationary operator $\hat{\sigma}_++\hat{\sigma}_-$, and its instantaneous eigenbasis identifies invariant-preserving trajectories while restricting the set of allowable control fields, as will be discussed in Sec.~\ref{sec:shortcuts}. The invariant can be written explicitly as
\begin{equation}\label{I_TLS_gen}
\begin{split}
    \hat{I}_{\texttt{TLS}}(t) &= \hat{S}_{\texttt{TLS}}(t)(\hat{\sigma}_++\hat{\sigma}_-)\hat{S}^{-1}_{\texttt{TLS}}(t),\\
    &=\frac{a \rho^2}{\theta_0}\hat{\sigma}_+ + \left[\frac{\theta_0}{a \rho^2} + \frac{\rho^2}{a \theta_0}\left(\frac{\dot{\rho}}{\rho} + \frac{\dot{a}}{2a} +ib\right)^2\right]\hat{\sigma}_-\\
    &-i  \frac{\rho^2}{\theta_0}\left(\frac{\dot{\rho}}{\rho} + \frac{\dot{a}}{2a} +ib\right)\hat{\sigma}_z.
\end{split}
\end{equation}
By construction, the relation $d\hat{I}_{\texttt{TLS}}/dt = \partial_t\hat{I}_{\texttt{TLS}} - i[\hat{I}_{\texttt{TLS}},\hat{H}_{\texttt{TLS}}] = 0$ is satisfied identically, establishing $\hat{I}_{\texttt{TLS}}(t)$ as a genuine dynamical invariant. Consequently, this operator encodes the nonadiabatic evolution paths of the driven two–level system and furnishes a framework for expressing the time–dependent control functions $a(t)$ and $b(t)$ in terms of its instantaneous eigenstructure~\cite{Leach_1977,Leach_1978}.

Once the temporal factorization leading to the scaled Hamiltonian $\hat{H}_{\mathbf{S}}(t)$, Eq.~\eqref{H_S}, has been obtained, the state vector is determined by the corresponding evolution operator. Since $[\hat{H}_{\mathbf{S}}(t_j),\hat{H}_{\mathbf{S}}(t_k)] = 0$ for any $t_j,t_k$, the Hamiltonian commutes with itself at all times. As a consequence, the time-ordering operator is trivial, and the Schr\"odinger equation in the scaled frame integrates directly to an ordinary exponential (no time ordering is needed). Transforming this solution back to the original frame gives:
\begin{equation}
\begin{split}\label{solucion_G}
    \ket{\psi(t)} &= \hat{S}_{\texttt{TLS}}(t)\exp\!\left[-i\theta_0 f(t)(\hat{\sigma}_++\hat{\sigma}_-)\right]\hat{S}_{\texttt{TLS}}^{-1}(t_0)\ket{\psi(t_0)},\\
    &=
    \begin{pmatrix}
        U_{11}(t)& U_{12}(t)\\  
        U_{21}(t)& U_{22}(t)
    \end{pmatrix}\ket{\psi(t_0)}
\end{split}
\end{equation}
where $f(t)=\int_{t_0}^t \mathrm{d}t'/\rho^2(t')$ represents the accumulated phase time in the scaled frame, and $\theta_0=\theta(t_0)$. The corresponding matrix elements of the evolution operator are:
\begin{subequations}
\begin{align}
    U_{11}(t) &= R \left[\frac{\cos(\theta_0 f(t))}{R_0} -\frac{\Omega_0R_0}{a_0} \sin(\theta_0 f(t)) \right], \label{Eq_U11} \\
    U_{12}(t) &= -i R_0R \sin(\theta_0 f(t)), \label{Eq_U12} \\
    U_{21}(t) &= i\left[\frac{\Omega R \cos(\theta_0 f(t))}{aR_0} - \frac{\sin(\theta_0 f(t))}{R_0R} -\frac{\Omega_0}{a_0} U_{22}(t)\right], \label{Eq_U21} \\
    U_{22}(t)  &= R_0\left[\frac{\cos(\theta_0 f(t))}{R} + \frac{\Omega R}{a} \sin(\theta_0 f(t))\right], \label{Eq_U22} 
\end{align}
\end{subequations}
with $R(t) = \sqrt{a(t)/\theta_0}\,\rho(t)$ and $\Omega(t) = \dot{a}/(2a) + \dot{\rho}/\rho + i b$. Quantities labeled with the subscript $0$ are evaluated at $t_0$. The terms $\Omega_0 R_0/a_0$ appearing in Eq.~\eqref{Eq_U11} originate from the non-unitary $\hat{G}$ transformation and have been validated through direct numerical integration of the Schr\"odinger equation~\cite{qutip5}. The state vector $\ket{\psi(t)}$ given in Eq.~\eqref{solucion_G} thus provides the general solution for the driven two–level system with coupling $a(t)$ and detuning $b(t)$, assuming that the Ermakov–Pinney equation for $\rho(t)$ admits a solution. It expresses the dynamics in terms of the auxiliary scaling function $\rho(t)$ and the accumulated scaled time $f(t)$, and it represents one of the central results of this work.

\section{Examples}\label{sec:examples}
\subsection{Landau–Zener-Majorana-Stückelberg transitions}
The Landau–Zener-Majorana-Stückelberg model~\cite{Landau_1932, Zener_1932, Stuckelberg_1932, Majorana_1932, Shevchenko_2010} characterizes non-adiabatic transitions in a driven two–level system passing through an avoided crossing. Within the framework introduced in the previous section, this setup corresponds to a linearly varying detuning $b(t) = \alpha t$ and a constant coupling $a(t) = g$. The dynamics reduce to a second-order differential equation closely related to the Weber equation, whose solutions in terms of parabolic cylinder functions capture the oscillatory corrections due to a finite interaction time~\cite{Vitanov_1996}. The associated Hamiltonian, obtained from Eq.~\eqref{H_TLS}, reads
\begin{equation}
    \hat{H}_{\texttt{LZMS}}(t) = g (\hat{\sigma}_+ + \hat{\sigma}_-) + \alpha t\, \hat{\sigma}_z.
\end{equation}
By inserting the linear detuning and constant coupling into the auxiliary condition defined in Eq.~\eqref{theta}, one obtains the specific complex Ermakov–Pinney equation relevant for this protocol. The resulting generalized frequency is
\begin{equation}
    \theta^2_{\texttt{LZMS}}(t) = g^2 + \alpha^2 t^2 + i\alpha,
\end{equation}
where the imaginary part $i\alpha$ originates from the time derivative of the detuning and encodes the non-adiabatic effects set by the sweep rate. The system’s evolution is then governed by the scaling function $\rho(t)$, which satisfies
\begin{equation}
    \ddot{\rho} + \theta^2_{\texttt{LZMS}}(t)\rho = \frac{\theta_{0, \texttt{LZMS}}^2}{\rho^3},
\end{equation}
with $\theta_{0, \texttt{LZMS}} = \theta_{\texttt{LZMS}}(t_0) = \sqrt{g^2+\alpha^2t_0^2+i\alpha}$. An important subtlety arises for the auxiliary Ermakov–Pinney equation when the generalized frequency $\theta^2(t)$ is complex: unlike the conventional time–dependent harmonic oscillator, where $\rho(t)$ remains real, here the auxiliary function $\rho(t)$ naturally becomes complex to reflect the non–Hermitian character of the intermediate description. The transition probabilities follow from the matrix elements given in Eqs.~\eqref{Eq_U11}--\eqref{Eq_U22}, using the scaling parameters
\begin{equation}
    R = \sqrt{\frac{g}{\theta_{0, \texttt{LZMS}}}}\rho(t),\quad \Omega = \frac{\dot{\rho}}{\rho} + i\alpha t.
\end{equation}

 \begin{figure}
    \centering
    \includegraphics[width=\linewidth]{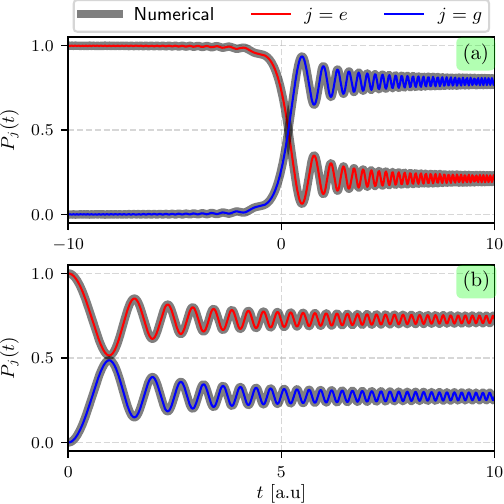}
    \caption{Time evolution of the excited-state population $P_e(t)$ and ground-state population $P_g(t)$ for the Landau–Zener–Majorana–Stückelberg protocol with $g=1$ and $\alpha=2g$. Panel (a) corresponds to the initial time $t_0=-10$, whereas panel (b) is obtained for $t_0=0$. The solid curves show the semi-analytical prediction from the invariant-based propagator [Eqs.~\eqref{Eq_U11}--\eqref{Eq_U22} evaluated with the Landau–Zener-Majorana-Stückelberg parameters]. The thick gray curves depict the numerical solution of the time–dependent Schrödinger equation. Both approaches capture the nonadiabatic dynamics, including the transient oscillations near the avoided crossing and the asymptotic Landau–Zener transition probability $\exp(-\pi g^2/\alpha)$.}
    \label{fig_lzms}
\end{figure}
We analyze two scenarios in which the system is initially prepared in the excited state, $\ket{\psi(t_0)} = \ket{e}$. In the first scenario, we choose $t_0=-10$ (dimensionless units), so that the system is prepared far from resonance, before the avoided crossing. In the second scenario, we set $t_0=0$, initializing the system exactly at resonance. For both choices of initial time, the state vector at a later time $t$ follows from the propagator matrix in Eqs.~\eqref{Eq_U11}--\eqref{Eq_U22} as
\begin{equation}
    \ket{\psi(t)} = \begin{pmatrix}U_{11}(t)\\U_{21}(t)\end{pmatrix}.
\end{equation}
The corresponding instantaneous probabilities are
\begin{equation}
    P_e(t) = |U_{11}(t)|^2, \qquad P_g(t) = |U_{21}(t)|^2,
\end{equation}
where $U_{11}(t)$ and $U_{21}(t)$ are entries of the propagator matrix, ensuring conservation of total probability, $P_e(t)+P_g(t)=1$. Figure~\ref{fig_lzms} presents a comparison between the analytical and numerical solutions. In panel (a), the system is initialized far from resonance ($t_0=-10$), capturing the transient oscillations and population behavior characteristic of the Landau–Zener–Majorana–Stückelberg model. Here, $P_e(t\to+\infty)$ denotes the probability of remaining in the excited state, while $P_g(t\to+\infty)$ is the probability of ending in the ground state, which asymptotically approaches $\exp(-\pi g^2/\alpha)$. Panel (b) illustrates the dynamics for the resonant initial condition ($t_0=0$). In this case, the initial frequency is $\theta_{0,\texttt{LZMS}}=\sqrt{g^2+i\alpha}$. This approach captures the evolution across the avoided crossing in both the adiabatic and diabatic regimes.

\begin{figure}
    \centering
    \includegraphics[width=\linewidth]{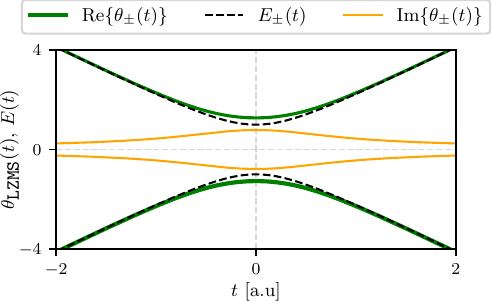}
    \caption{Time evolution of the energy levels and of the non-adiabatic complex frequency in the Landau–Zener–Majorana–Stückelberg model for an avoided level crossing. The detuning $b(t) = \alpha t$ is linearly swept through resonance at $t=0$ with a constant coupling $g$. The adiabatic eigenenergies $E_{\pm}(t) = \pm \sqrt{g^2 + \alpha^2 t^2}$ exhibit an avoided crossing with a minimum gap of $2g$. The system’s non-adiabatic dynamics are described by the complex frequency $\theta_{\texttt{LZMS}}(t)$, whose real part is determined by the adiabatic potential $E^2(t)$, while the imaginary contribution $i\alpha$ accounts for the non-adiabatic effects arising from the finite sweep rate~\cite{Vitanov_1996}.}
    \label{fig_2}
\end{figure}
The frequency $\theta_{\texttt{LZMS}}(t)$ links the system’s physical parameters to the non-adiabatic dynamics. With the adiabatic energies defined as $E(t) = \sqrt{g^2 + \alpha^2 t^2}$, the complex frequency can be written as $\theta^2_{\texttt{LZMS}}(t) = E^2(t) + i\alpha$. In this expression, the real part is given by the squared adiabatic energy $E^2(t)$, while the imaginary contribution $i\alpha$ encodes the sweep rate, as illustrated in Fig.~\ref{fig_2}. Far from resonance ($t \to \pm\infty$), $E^2(t)$ exceeds $\alpha$, and the magnitude of $\theta_{\texttt{LZMS}}(t)$ converges to the adiabatic energy value. Near the avoided crossing ($t \approx 0$), the imaginary contribution becomes most significant relative to the level splitting $g^2$. 

\subsection{Non–Hermitian Landau–Zener–Majorana–Stückelberg transitions}
Non–Hermitian Hamiltonians serve as effective models for open quantum systems, capturing dissipation, decay, amplification, and parity–time (PT) symmetric behavior~\cite{Moiseyev_2011,Bender_2007,Torosov_2017}. Here, we investigate a particular non–Hermitian extension of the Landau–Zener–Majorana–Stückelberg model within the unified algebraic framework. The detuning is chosen to be purely imaginary and linear in time, $b(t) = i\alpha t$, describing a sweep through a dissipative avoided crossing, while the coupling is held constant, $a(t) = g$. The Hamiltonian governing this dissipative dynamics is
\begin{equation}
    \hat{H}_{\texttt{NHLZMS}}(t) =  g(\hat{\sigma}_+ + \hat{\sigma}_-) + i\alpha t\,\hat{\sigma}_z .
\end{equation}
Inserting the linear imaginary detuning and time-independent coupling into the generalized auxiliary condition leads to the specific Ermakov–Pinney equation associated with this non–Hermitian model. The resulting effective frequency $\theta^2_{\texttt{NHLZMS}}(t)$ takes the form
\begin{equation}
    \theta^2_{\texttt{NHLZMS}}(t) = g^2 - \alpha^2 t^2 - \alpha,
\end{equation}
with the constant reference frequency defined as $\theta_{0, \texttt{NHLZMS}} = \theta_{\texttt{NHLZMS}}(t_0) = \sqrt{g^2-\alpha^2t_0^2-\alpha}$. The additional constant term $-\alpha$ arises from the time derivative of the imaginary detuning. The corresponding scaling parameters for this dissipative evolution are
\begin{equation}\label{ROmega_nhlzms}
    R = \sqrt{\frac{g}{\theta_{0, \texttt{NHLZMS}}}}\rho(t),\quad \Omega = \frac{\dot{\rho}}{\rho} - \alpha t.
\end{equation}
In contrast to the standard Landau–Zener–Majorana–Stückelberg scenario, where $\theta^2_{\texttt{LZMS}}(t)$ is complex, here $\theta^2_{\texttt{NHLZMS}}(t)$ is strictly real and forms an inverted parabolic profile $-\alpha^2 t^2$ in time.
\begin{figure}
    \centering
    \includegraphics[width=\linewidth]{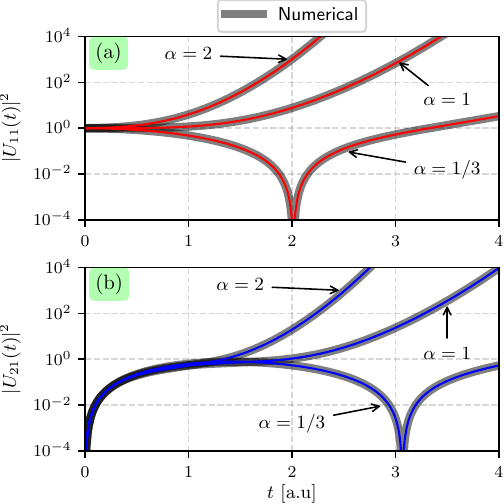}
    \caption{Time evolution of the squared moduli of the propagator elements $|U_{11}(t)|^2$ and $|U_{21}(t)|^2$ for the non–Hermitian Landau–Zener–Majorana–Stückelberg model with imaginary detuning $b(t) = i\alpha t$, assuming the system is initially in the excited state. In contrast with the Hermitian case, the total norm is not conserved ($|U_{11}(t)|^2 + |U_{21}(t)|^2 \neq 1$) because the dynamics are non-unitary. The evolution exhibits an initial oscillatory transient near the avoided crossing, followed by a marked exponential increase of the squared amplitudes once the system crosses the dynamical exceptional point and enters the broken-symmetry regime. This exponential amplification is a direct manifestation of the state-dependent gain and loss introduced by the anti-Hermitian part of the Hamiltonian.}
    \label{fig_3}
\end{figure}
This frequency profile is thus analogous to a scattering problem over a potential barrier, rather than motion within a confining potential well. The deviation from unitarity stems from the anti-Hermitian contribution $i\alpha t \hat{\sigma}_z$. Throughout the evolution, the Lewis–Ermakov invariant for this two–level system remains conserved. The exact state vector follows from the propagator elements $U_{11}(t)$ and $U_{21}(t)$ under the initial excited–state condition [Eqs.~\eqref{Eq_U11}--\eqref{Eq_U21}], evaluated using the scaling functions $R$ and $\Omega$. Their squared magnitudes as functions of time are presented in Fig.~\ref{fig_3}. In time–dependent non–Hermitian systems, the onset of dynamical symmetry breaking generally does not coincide with the threshold inferred from the instantaneous eigenvalue spectrum.

\begin{figure}
    \centering
    \includegraphics[width=\linewidth]{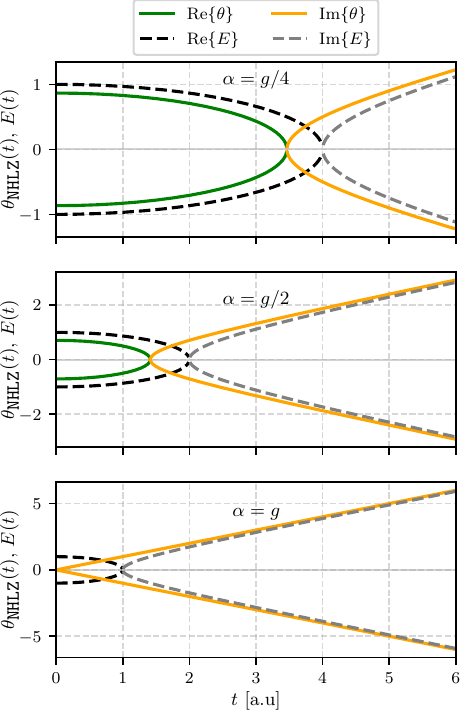}
    \caption{Time evolution of the non–Hermitian Landau–Zener–Majorana–Stückelberg model for different sweep rates $\alpha \in \{g/4, g/2, g\}$ with $g=1$. Solid and dashed curves show the real and imaginary parts of the adiabatic energies $E(t)$, respectively, while the associated effective frequency profiles are given by $\theta^2_{\texttt{NHLZMS}}(t) = g^2 - (\alpha t)^2 - \alpha$. Solid vertical lines denote the static exceptional points ($t = g/\alpha$), and dashed vertical lines indicate the dynamical exceptional points ($t = \sqrt{g^2-\alpha}/\alpha$). The comparison demonstrates that larger sweep rates $\alpha$ increase the delay between the static onset of symmetry breaking and the actual emergence of exponential amplification.}
    \label{fig_4}
\end{figure}

For the two–level system described by $\hat{H}_{\texttt{NHLZMS}}(t)$, the instantaneous eigenvalues are given by $E(t) = \pm\sqrt{g^2 - (\alpha t)^2}$. The static exceptional point---at which the adiabatic branches coalesce and pseudo-Hermiticity breaks down~\cite{Heiss_2012}---follows from the condition $E(t) = 0$, leading to $t = g/\alpha$. This characterization relies on a strictly adiabatic viewpoint and neglects non-adiabatic effects stemming from the finite sweep rate $\alpha$. By contrast, the actual time evolution is dictated by the auxiliary equation in Eq.~\eqref{Ermakov_TLS}, where the effective frequency $\theta^2_{\texttt{NHLZMS}}(t) = g^2 - (\alpha t)^2 - \alpha$ determines the dynamical stability. The additional term $-\alpha$ arises from the time dependence of the Hamiltonian and plays the role of an inertial contribution induced by the parameter sweep. As shown in Fig.~\ref{fig_4}, the transition from oscillatory motion to exponential amplification does not align with the static exceptional point. Instead, the dynamical exceptional point is set by the zero of the effective frequency, $\theta^2_{\texttt{NHLZMS}}(t) = g^2 - (\alpha t)^2 - \alpha = 0$. This delayed onset of symmetry breaking is a direct consequence of the finite ramp speed, which shifts the critical boundary and preserves oscillatory dynamics beyond the threshold predicted by the instantaneous Hamiltonian.

\subsection{Adiabatic rapid passage}
Adiabatic rapid passage (ARP) is a powerful and reliable method for achieving complete population transfer between quantum states, and it is widely applied in nuclear magnetic resonance, quantum optics, and quantum information processing~\cite{Vitanov_2001,Vitanov_2001A}. The mechanism behind population inversion can be described as adiabatic following of the instantaneous dressed states: during the frequency sweep, the system remains in a single instantaneous eigenstate and thereby evolves smoothly from one bare state to the other~\cite{Camparo_1984,Berry_1984}. In this scheme, the detuning is chirped as $b(t) = \Delta_0 \tanh(t/\tau)$, sweeping continuously from a large negative value at $t \to -\infty$ to a large positive value at $t \to +\infty$, while the coupling is shaped as $a(t) = g\,\text{sech}(t/\tau)$, peaking at the resonance crossing $t=0$. Here, $\tau$ sets the characteristic pulse duration and $\Delta_0$ fixes the amplitude of the detuning. The corresponding Hamiltonian reads
\begin{equation}
    \hat{H}_{\texttt{ARP}} = \Delta_0 \tanh(t/\tau)\,\hat{\sigma}_z + g\,\text{sech}(t/\tau)(\hat{\sigma}_+ + \hat{\sigma}_-).
\end{equation}
The combination of a hyperbolic-secant pulse envelope with a hyperbolic tangent frequency chirp realizes the exactly solvable Allen–Eberly model, which serves as a standard example of adiabatic rapid passage and coherent population inversion in two–level systems~\cite{Allen_1975}.

Inserting these time dependences into the general auxiliary condition Eq.~\eqref{theta}, the effective complex frequency appearing in the Ermakov–Pinney equation becomes
\begin{equation}
    \begin{split}
         \theta^2_{\texttt{ARP}}(t) &= g^2\text{sech}^2(t/\tau) + \Delta_0^2\tanh^2(t/\tau) + i\frac{\Delta_0}{\tau} 
         \\
    &- \frac{1}{2\tau^2} + \frac{1}{4\tau^2}\tanh^2(t/\tau).
    \end{split}
\end{equation}
The initial frequency parameter $\theta_{0, \texttt{ARP}} = \sqrt{g^2 + i\Delta_0/\tau - 1/(2\tau^2)}$ guarantees that the auxiliary scaling function $\rho(t)$ solves the complex Ermakov–Pinney equation Eq.~\eqref{Ermakov_TLS} with the specified initial conditions $\rho(t_0)=1$ and $\dot{\rho}(t_0) = 0$. The associated scaling parameters entering the exact propagator Eqs.~(\ref{Eq_U11})–(\ref{Eq_U22}) are then
\begin{equation}
\begin{split}
    R &= \sqrt{\frac{g\text{sech}(t/\tau)}{\theta_{0, \texttt{ARP}}}}\rho(t),\\
    \Omega &= -\frac{1}{2\tau}\tanh(t/\tau) + \frac{\dot{\rho}}{\rho} + i\Delta_0 \tanh(t/\tau).
\end{split}
\end{equation}

\begin{figure}
    \centering
    \includegraphics[width=\linewidth]{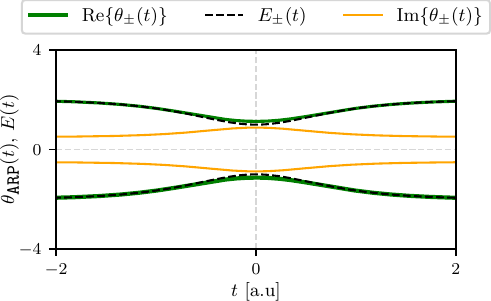}
    \caption{Time evolution of the adiabatic energies $E_{\pm}(t)$ and of the non-adiabatic complex frequency $\theta_{\texttt{ARP}}(t)$ for the adiabatic rapid passage protocol. The coupling $a(t)=g\text{sech}(t/\tau)$ and detuning $b(t)=\Delta_0\tanh(t/\tau)$ generate a smooth avoided crossing at $t=0$ with a minimum gap of $2g$. The real part of $\theta_{\texttt{ARP}}(t)$ tracks the adiabatic energy $E_+(t)$ away from resonance and departs from it near the avoided crossing, while the imaginary part of $\theta_{\texttt{ARP}}(t)$ captures the finite sweep-rate non-adiabaticity of the chirped drive, directly analogous to the imaginary contribution $i\alpha$ in the Landau–Zener–Majorana–Stückelberg scenario (Fig.~\ref{fig_2}).}
    \label{fig_6}
\end{figure}

The adiabatic eigenenergies of the ARP Hamiltonian are $E_{\pm}(t) = \pm \sqrt{g^2 \operatorname{sech}^2(t/\tau) + \Delta_0^2 \tanh^2(t/\tau)}$, which exhibit a smooth avoided crossing at $t=0$ with a minimum gap of $2g$, set by the peak Rabi frequency. The nonadiabatic frequency $\theta_{\texttt{ARP}}$ acquires an imaginary contribution $i\Delta_0/\tau - 1/(2\tau^2) + \tanh^2(t/\tau)/(4\tau^2)$, capturing the sweep-induced deviations from perfect adiabaticity, in direct correspondence with the Landau–Zener–Majorana–Stückelberg framework. This is visualized in Fig.~\ref{fig_6}: the real component of $\theta_{\texttt{ARP}}$ tracks the adiabatic energy surfaces, while the imaginary component encodes corrections due to the finite sweep rate.

The power of the invariant-based method becomes particularly evident when analyzing population transfer under ARP. The adiabaticity parameter $\mathcal{A}=g\tau$ sets the fidelity of the transfer: for $\mathcal{A}\gg 1$, the system closely follows the instantaneous eigenstate of $\hat{H}_{\texttt{ARP}}$ and nearly complete inversion is achieved, whereas for $\mathcal{A}\lesssim 1$ nonadiabatic contributions generate residual oscillations and hinder full transfer. The propagator derived from Eqs.~(\ref{Eq_U11})--(\ref{Eq_U22}), in combination with the exact solution $\rho(t)$ of Eq.~(\ref{Ermakov_TLS}), provides an accurate, approximation-free description in both limits. For an initially excited state $|\psi(t_0)\rangle = |e\rangle$, the instantaneous level populations are
\begin{equation}
    P_e(t) = |U_{11}(t)|^2, \qquad P_g(t) = |U_{21}(t)|^2,
\end{equation}
with $P_e(t) + P_g(t) = 1$ remaining constant under the unitary evolution.

\begin{figure}
    \centering
    \includegraphics[width=\linewidth]{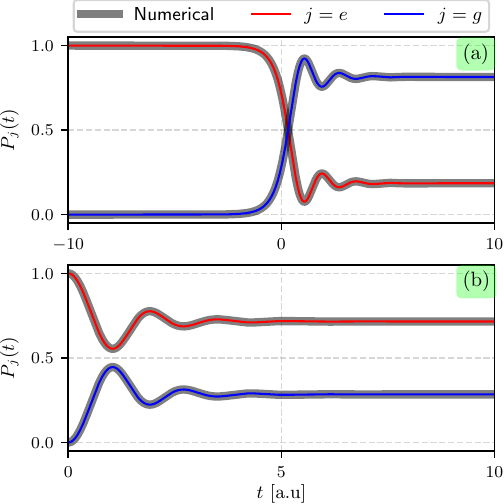}
    \caption{Time evolution of the excited-state population $P_e(t)$ and ground-state population $P_g(t)$ for the adiabatic rapid passage protocol with parameters $g=\tau=1$, $\Delta_0=2g$. Panel (a) depicts the dynamics for an initial time $t_0=-10$, where the system is initialized far from the resonance, while panel (b) corresponds to initialization at the pulse maximum $t_0=0$. Solid lines represent the semi-analytical solution, and thick gray lines show the numerical integration of the time–dependent Schrödinger equation.}
    \label{fig_5}
\end{figure}

Figure~\ref{fig_5} contrasts these semi-analytical predictions with direct numerical solutions of the time–dependent Schrödinger equation for two distinct choices of the initial time. Panel (a) illustrates the evolution for $t_0=-10$, where the system is prepared long before it passes through resonance, whereas panel (b) shows the case $t_0=0$, with the system initialized at the peak of the coupling pulse. For both initial conditions, the semi-analytical curves are visually identical to the numerical results, confirming that the Lewis–Ermakov construction is exact for this type of driving. In the adiabatic regime ($g\tau\gg 1$), population transfer occurs smoothly through the avoided crossing and is strongly influenced by the imaginary part of $\theta_{\texttt{ARP}}(t)$. Experimental implementations include coherent population transfer in trapped ions~\cite{An_2016}, stimulated Raman adiabatic passage in ultracold molecules~\cite{Vitanov_2017}, and robust qubit gates in superconducting circuits~\cite{Vepsalainen_2019}, all exhibiting high fidelity and robustness to parameter variations.

\section{Shortcuts to adiabaticity in two–level systems}\label{sec:shortcuts}
The previous examples examined forward dynamics driven by prescribed fields $a(t)$ and $b(t)$. Shortcuts to adiabaticity~\cite{Chen_2010, Torrontegui_2013, GueryOdelin_2019} reverse this perspective: the objective becomes to engineer $a(t)$, $b(t)$, and the scaling factor $\rho(t)$ so that the system follows a specified transitionless trajectory. The dynamical invariant $\hat{I}_{\texttt{TLS}}(t)$ [Eq.~\eqref{I_TLS_gen}] is constructed to satisfy $d\hat{I}_{\texttt{TLS}}/dt = 0$ and, in line with Lewis–Riesenfeld theory~\cite{Lewis_1969,Ramos_2018}, provides the central theoretical tool of this framework. Yet, the mere conservation of this invariant does not ensure that a system initially prepared in an energy eigenstate will evolve into a desired final energy eigenstate without residual excitations. Achieving high-fidelity transitionless control requires an extra boundary condition: at the protocol endpoints $t_b \in \{t_0, t_f\}$, the invariant must commute with the Hamiltonian, so that both operators share the same eigenbasis:
\begin{equation}\label{Eq_comm_boundary}
    \bigl[\hat{I}_{\texttt{TLS}}(t_b),\,\hat{H}_{\texttt{TLS}}(t_b)\bigr] = 0.
\end{equation}
When this commutation relation is satisfied, the eigenstates of the invariant align with the instantaneous energy eigenstates at the initial and final times, thereby suppressing quantum friction---i.e., the unwanted excitations arising from non-adiabatic evolution---in the framework of transitionless quantum driving~\cite{Berry_2009}.

In the standard formulation of dynamical invariant theory, the integration constant $\theta_0$ is fixed by the initial boundary conditions of the physical Hamiltonian. In the context of quantum control and shortcuts to adiabaticity, however, this viewpoint can be reversed. If we promote $\vartheta_0$ from a fixed initial boundary parameter to a free geometric parameter, we gain an additional degree of freedom. Optimizing over this parameter enables us to tailor the auxiliary boundary conditions of $\rho(t)$ at $t_0$ and $t_f$ while keeping the actual physical boundary states of the qubit unchanged. Thus, $\vartheta_0$ functions as a geometric control knob that can be tuned to reduce transient energy excitations or suppress unwanted nonadiabatic transitions in fast driving protocols.

To clarify the algebra embedded in Eq.~\eqref{Eq_comm_boundary}, we explicitly compute the commutator at the boundaries. For forward (natural) dynamics, the invariant is usually built with a fixed Ermakov constant determined by the initial frequency, $\theta_0 = \theta(t_0)$, which in turn enforces the normalization $\rho(t_0) = 1$~\cite{Lewis_1969}. This strict constraint overconstrains the inverse-engineering task, making it impossible for the scaling function to meet the commutation requirements at both $t_0$ and $t_f$ simultaneously. To overcome this issue, we formally extend the invariant $\hat{I}_{\texttt{TLS}}(t)$ and the complex Ermakov–Pinney equation by replacing $\theta_0$ with a free geometric constant $\vartheta_0$.
Assuming, as usual, that the control fields and the scaling function are stationary at the boundaries ($\dot{a}(t_b) = \dot{\rho}(t_b) = \ddot{\rho}(t_b) = 0$), a direct evaluation of the commutator at $t_b$ yields
\begin{equation}
    \bigl[\hat{I}_{\texttt{TLS}}(t_b),\,\hat{H}_{\texttt{TLS}}(t_b)\bigr]= \frac{\rho^2(t_b)}{\vartheta_0}\,\gamma(t_b)\left(\hat{\sigma}_z-\frac{2\,b(t_b)}{a(t_b)}\hat{\sigma}_-\right),
\end{equation}
where $\rho(t)$ satisfies the generalized complex Ermakov–Pinney equation $\ddot{\rho}+\theta^2(t)\rho=\vartheta_0^2/\rho^3$ (with $\theta^2(t)$ given in Eq.~\eqref{theta}), and
\begin{equation}
    \gamma(t_b)= a^2(t_b)+b^2(t_b)-\frac{\vartheta_0^2}{\rho^{4}(t_b)}
\end{equation}
quantifies the deviation between the invariant basis and the energy eigenbasis. Consequently, the commutator is zero if and only if $\gamma(t_b)=0$, which determines the boundary values of the scaling function:
\begin{equation}\label{Eq_rho_boundary}
    \rho(t_b) = \left(\frac{\vartheta_0^2}{a^2(t_b)+b^2(t_b)}\right)^{1/4}.
\end{equation}

The function of $\vartheta_0$ is thus clarified: by decoupling the Ermakov constant from the system’s instantaneous initial energy, it supplies precisely the mathematical flexibility required for $\rho(t)$ to interpolate smoothly between the stationarity constraints at $t_0$ and $t_f$. In particular, one may choose $\vartheta_0^2 = a^2(t_0) + b^2(t_0)$ so that it coincides with the initial physical parameters, thereby normalizing the scaling function at the beginning of the protocol, $\rho(t_0) = 1$. Under this choice, the target boundary condition at the final time $t_f$ is uniquely determined by the ratio of the initial and final adiabatic energy splittings:
\begin{equation}
    \rho(t_f) = \left(\frac{a^2(t_0) + b^2(t_0)}{a^2(t_f) + b^2(t_f)}\right)^{1/4}.
\end{equation}

Once the boundary values are established by Eq.~\eqref{Eq_rho_boundary}, the next step is to specify a smooth phenomenological trajectory $\rho(t)$ that seamlessly interpolates between them while respecting the boundary stationarity constraints $\dot{\rho}(t_b)=0$ and $\ddot{\rho}(t_b)=0$. A standard and mathematically robust choice is a quintic polynomial ansatz~\cite{Chen_2010, Torrontegui_2013} parameterized by the normalized time $s = (t - t_0)/(t_f - t_0)$:
\begin{equation} \label{Eq_rho_ansatz}
    \rho(s) = \rho(t_0) + \bigl[\rho(t_f) - \rho(t_0)\bigr]\bigl(10s^3 - 15s^4 + 6s^5\bigr).
\end{equation}
This ansatz defines a $\mathcal{C}^2$-continuous trajectory in the complex plane, structurally guaranteeing that the boundary commutators in Eq.~\eqref{Eq_comm_boundary} are exactly satisfied.

With the polynomial trajectory $\rho(t)$ prescribed, the reverse-engineering of the requisite control fields is driven by the generalized complex Ermakov–Pinney equation. Because the physical parameters $a(t)$ and $b(t)$ are not fixed in advance in the control scheme, the effective complex frequency $\theta^2(t)$ is extracted directly from the prescribed trajectory as $\theta^2(t) = \vartheta_0^2/\rho^4(t) - \ddot{\rho}(t)/\rho(t)$. Equating this constructed frequency to the theoretical expression derived in Eq.~\eqref{theta} generates a system of two coupled real differential equations, effectively splitting the real and imaginary components:
\begin{subequations}
\begin{align}
    \text{Re}\left[\frac{\vartheta_0^2}{\rho^4} - \frac{\ddot{\rho}}{\rho}\right] &= a^2 + b^2 + \frac{\ddot{a}}{2a} - \frac{3\dot{a}^2}{4a^2},\\
    \text{Im}\left[\frac{\vartheta_0^2}{\rho^4} - \frac{\ddot{\rho}}{\rho}\right] &= \dot{b} - \frac{\dot{a}b}{a}.
\end{align}
\end{subequations}
Because this framework provides two independent conditions for the two unknown physical fields $a(t)$ and $b(t)$, the control system is exactly determined. Consequently, these equations can be directly inverted to yield both driving protocols simultaneously, entirely eliminating the need to impose arbitrary or phenomenological constraints on either field. The resulting reverse-engineered fields guarantee that the system evolves exactly along the target non-adiabatic path dictated by the scaling function $\rho(t)$.

To resolve this system in its most general form, we first address the scenario where the scaling function $\rho(t)$ is complex, meaning the imaginary component in the left-hand side of Eq.~\eqref{theta} is non-zero, $F_I(t) = \text{Im}[\vartheta_0^2/\rho^4 - \ddot{\rho}/\rho] \neq 0$. In this case, the coupled differential equations can still be systematically reduced. The imaginary condition $\dot{b} - \dot{a}b/a = F_I(t)$ is a first-order linear ordinary differential equation for $b(t)$, which can be integrated directly to express the detuning in terms of the coupling. By introducing the angle $\phi$ defined by the ratio of the initial values of the driving fields, $\tan\phi = b_0/a_0$, where $a_0 = a(t_0)$ and $b_0 = b(t_0)$, the solution reads:
\begin{equation}\label{Eq_general_b_solution}
    b(t) = a(t) \left[ \tan\phi + \int_{t_0}^t \frac{F_I(t')}{a(t')} dt' \right].
\end{equation}
By defining the Ermakov variable $y(t) = a(t)^{-1/2}$ and substituting Eq.~\eqref{Eq_general_b_solution} into the real part of the Ermakov frequency, we obtain a highly non-linear integro-differential equation for $y(t)$:
\begin{equation}\label{Eq_integro_differential}
    \ddot{y}(t) + F_R(t) y(t) = \frac{1 + \left[ \tan\phi + \int_{t_0}^t F_I(t') y(t')^2 dt' \right]^2}{y(t)^3},
\end{equation}
where $F_R(t) = \text{Re}[\vartheta_0^2/\rho^4 - \ddot{\rho}/\rho]$. Solving Eq.~\eqref{Eq_integro_differential} for $y(t)$ yields $a(t) = y(t)^{-2}$, which in turn determines $b(t)$ via Eq.~\eqref{Eq_general_b_solution}. Because of the integral term in the numerator, Eq.~\eqref{Eq_integro_differential} generally does not admit closed-form analytical solutions for arbitrary scaling functions. It can be integrated numerically, and developing perturbative or analytical approximations for it remains an open challenge.

A physically motivated simplification arises when $\rho(t)$ is restricted to be purely real. Under this restriction, the imaginary component $F_I(t)$ vanishes identically, and the general solution in Eq.~\eqref{Eq_general_b_solution} collapses to a simple proportionality between the physical fields, $b(t) = a(t) \tan\phi$. Physically, the angle $\phi$ represents the fixed orientation of the driving field vector in the Bloch sphere. Under this parametrization, the sum of the squares of the fields simplifies to $a^2 + b^2 = a^2 \sec^2\phi$. Substituting these relations into the real component of the Ermakov–Pinney equation leads to:
\begin{equation}\label{Eq_real_a}
    \frac{\vartheta_0^2}{\rho^4} - \frac{\ddot{\rho}}{\rho} = a^2\sec^2\phi + \frac{\ddot{a}}{2a} - \frac{3\dot{a}^2}{4a^2}.
\end{equation}
By executing the standard Ermakov substitution $y(t) = a(t)^{-1/2}$, Eq.~\eqref{Eq_real_a} maps exactly onto the classical autonomous Ermakov–Pinney equation:
\begin{equation}
    \ddot{y} + \left(\frac{\vartheta_0^2}{\rho^4} - \frac{\ddot{\rho}}{\rho}\right) y = \frac{\sec^2\phi}{y^3}.
\end{equation}
Because $\rho(t)$ is strictly real, this equation admits the exact, explicit particular solution $y(t) = \sqrt{\sec\phi}\,\rho(t)/\sqrt{\vartheta_0}$. Returning to the original physical variables, we obtain the driving fields in a closed trigonometric form:
\begin{subequations}\label{Eq_explicit_fields}
\begin{align}
    a(t) &= \frac{\vartheta_0 \cos\phi}{\rho^2(t)},\\
    b(t) &= \frac{\vartheta_0 \sin\phi}{\rho^2(t)}.
\end{align}
\end{subequations}
The standard resonant shortcut protocol ($b(t)=0$) is immediately recovered from Eq.~\eqref{Eq_explicit_fields} by setting $\phi=0$. This reveals a mathematical and physical trade-off: enforcing a purely real scaling function $\rho(t)$ guarantees a fully explicit analytical determination of the control fields in terms of the orientation angle $\phi$, but constrains the driving protocol to proportional modulations where the detuning and coupling share the same temporal shape. Conversely, a complex scaling function $\rho(t)$ unlocks the freedom to design independent temporal profiles for $a(t)$ and $b(t)$, at the cost of having to solve the coupled non-linear equations in Eq.~\eqref{theta} numerically.

\begin{figure}
    \centering
    \includegraphics[width=\linewidth]{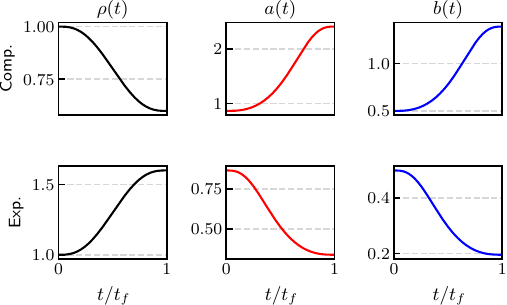}
    \caption{Representative shortcut to adiabaticity protocols obtained from the invariant-based inverse-engineering procedure for a real scaling function $\rho(t)$ with parameters $\vartheta_0=1$, $\phi = \pi/6$, $t_f=10$ and $\rho(t_0)=1$. The top row (Comp.) corresponds to a compression protocol with $\rho(t_f)=0.6$, while the bottom row (Exp.) corresponds to an expansion protocol with $\rho(t_f)=1.6$. Columns show, from left to right: the prescribed quintic polynomial trajectory $\rho(t)$ (Eq.~\eqref{Eq_rho_ansatz}), the reverse-engineered coupling $a(t)$ and detuning $b(t)$ field (Eq.~\eqref{Eq_explicit_fields}). Because the ratio $b(t)/a(t)=\tan\phi$ is constant, both fields share the same temporal profile and differ only by a fixed scaling factor.}
    \label{fig_7}
\end{figure}

To illustrate the inverse-engineering procedure in the analytically solvable real-$\rho$ regime, Fig.~\ref{fig_7} presents two representative shortcut protocols corresponding to compression and expansion processes.  
The upper row corresponds to a compression protocol, where the scaling function decreases monotonically from $\rho(t_0)$ to $\rho(t_f) < \rho(t_0)$, while the lower row corresponds to an expansion protocol with $\rho(t_f) >\rho(t_0)$. In both cases, the scaling function follows the quintic polynomial trajectory of Eq.~\eqref{Eq_rho_ansatz}, which guarantees continuity and exact satisfaction of the boundary conditions. The resulting coupling $a(t)$ and detuning $b(t)$, obtained from Eq.~\eqref{Eq_explicit_fields}, inherit the smooth temporal profile of $\rho(t)$ and remain proportional throughout the protocol. An important feature of the real-$\rho$ solution is that the ratio ($b(t)/a(t) = \tan \phi$) remains constant throughout the protocol. Therefore, the direction of the control vector on the Bloch sphere is preserved while only its magnitude is modulated. The shortcut is thus completely determined by the prescribed scaling trajectory $\rho(t)$, which acts as an effective control parameter governing the temporal compression or expansion of the energy landscape.

\section{Conclusions}\label{conclusions}
The main technical and conceptual contribution of this work is the explicit construction and practical use of the Lewis–Ermakov dynamical invariant for time–dependent two–level systems. The invariant,
\begin{equation*}
    \hat{I}_{\texttt{TLS}}(t) = \hat{S}_{\texttt{TLS}}(t)\bigl(\hat{\sigma}_+ + \hat{\sigma}_-\bigr)\hat{S}^{-1}_{\texttt{TLS}}(t),
\end{equation*}
given in closed form in Eq.~\eqref{I_TLS_gen}, fulfills $d\hat{I}_{\texttt{TLS}}/dt=0$ and thereby defines a time-independent reference frame for the driven qubit. In practice, this invariant encodes the system’s non-adiabatic trajectories, specifies the instantaneous invariant eigenbasis, and reduces the full time–dependent dynamics to a mere scalar phase accumulation in the scaled frame (Eqs.~\eqref{solucion_G} and \eqref{Eq_U11}--\eqref{Eq_U22}).

The invariant serves two complementary purposes. As a computational device, once $\rho(t)$—the solution to the complex Ermakov–Pinney equation, Eq.~\eqref{Ermakov_TLS}—is known, its combination with the invariant provides the exact propagator and wavefunction for arbitrary $a(t)$ and $b(t)$, covering also non-unitary regimes. This replaces problem-specific integrations with a single auxiliary equation. As a control resource, the invariant generates shortcuts to adiabaticity: enforcing $[\hat{I}_{\texttt{TLS}}(t_b),\hat{H}_{\texttt{TLS}}(t_b)]=0$ at the initial and final times fixes $\rho(t_b)$ via Eq.~\eqref{Eq_rho_boundary} and transforms any chosen invariant-preserving path into explicit control fields through Eqs.~\eqref{Ermakov_TLS}, \eqref{theta} and \eqref{Eq_explicit_fields}.

This twofold function—analysis and design—elevates the Lewis–Ermakov invariant to a central result of the manuscript. It provides the conceptual link that unifies continuous-variable (oscillator) and discrete-variable (two–level) realizations of $sl(2,\mathbb{C})$: the same sequence (factorization, Ermakov auxiliary equation, invariant construction, and propagator) applies in both settings, with the invariant delivering the operative shortcuts used in inverse-engineering approaches. The case studies considered here—the Hermitian and non–Hermitian Landau–Zener models and the Allen–Eberly implementation of adiabatic rapid passage—show that the invariant-based formulation recovers known non-adiabatic effects while extending control into domains where the instantaneous spectrum alone is not sufficient.

Adopting this invariant-centered viewpoint also leads to concrete experimental guidelines: first design a smooth $\rho(t)$ that satisfies the boundary conditions, then use the associated Ermakov equation to determine $\theta^2(t)$, and finally invert Eqs.~\eqref{theta} to obtain the physical control fields $a(t)$ and $b(t)$. When $\rho(t)$ is chosen to be real, the resulting protocols become fully explicit (Eq.~\eqref{Eq_explicit_fields}), enabling fast, reflectionless state preparation in finite time. Generalizations to higher-spin systems and hybrid architectures emerge from the same algebraic mapping and naturally retain the same invariant-based shortcut strategies.

Finally, let us consider the generalized time–dependent Hamiltonian for an $N$-level system, expressed as $\hat{H}_{\texttt{NLS}}(t) = a(t) \left( \hat{J}_+ + \hat{J}_- \right) + b(t) \hat{J}_z$, where $\hat{J}_{\pm, z}$ denote the standard angular momentum operators. By utilizing the fundamental commutation relations, namely $[\hat{J}_z, \hat{J}_\pm] = \pm \hat{J}_{\pm}$, and $[\hat{J}_+, \hat{J}_-] = 2 \hat{J}_z$,it is possible to extend the algebraic framework developed throughout this work. Consequently, one can derive a generalized dynamical invariant structured identically to the one presented in Eq.~\eqref{I_TLS_gen}. This formulation effectively extends the Lewis–Riesenfeld invariant approach from a simple two–level system to an arbitrary $N$-level system spanning a higher-dimensional Hilbert space. Remarkably, the overall algebraic structure remains unchanged: the dynamics are still governed by an auxiliary scaling function $\rho(t)$ satisfying an Ermakov–Pinney equation of the same form as Eq.~\eqref{Ermakov_TLS}, but with the effective frequency $\theta^2(t) = a^2 + b^2/4 + i\dot{b}/2 - i\dot{a}b/(2a)
+ \ddot{a}/(2a) - 3\dot{a}^2/(4a^2)$. Compared with the two–level case, the modifications appear only in the coefficients of the terms involving the detuning $b(t)$, reflecting the different normalization of the angular momentum generators in the general spin-$j$ representation. The corresponding dynamical invariant can be written explicitly as

\begin{equation}\label{I_NLS_gen}
\begin{split}
    \hat{I}_{\texttt{NLS}}(t) &= \frac{a \rho^2}{\theta_0}\hat{J}_+ + \left[\frac{\theta_0}{a \rho^2} + \frac{\rho^2}{a \theta_0}\left(\frac{\dot{\rho}}{\rho} + \frac{\dot{a}}{2a} +i\frac{b}{2}\right)^2\right]\hat{J}_--\\
    &-2i  \frac{\rho^2}{\theta_0}\left(\frac{\dot{\rho}}{\rho} + \frac{\dot{a}}{2a} +i\frac{b}{2}\right)\hat{J}_z.
\end{split}
\end{equation}
Therefore, the invariant-based approach presented in this work extends naturally from two–level systems to arbitrary $N$-level realizations of the $sl(2,\mathbb{C})$ algebra. The persistence of the same Ermakov–Pinney structure highlights the universality of the Lewis–Riesenfeld formalism and suggests a direct framework for the analysis and inverse engineering of multilevel quantum dynamics.
\appendix
\section{Lewis–Ermakov invariant for the harmonic oscillator}\label{A_A}
This appendix provides the mathematical foundation for the Lewis–Ermakov invariant by considering the general classical quadratic Hamiltonian, establishing the phase space geometry that guides the quantum realization in Appendix~\ref{A_B}. Following the framework of Refs.~\cite{Leach_1977, Leach_1978}, we seek a time–dependent canonical transformation that maps the system to an autonomous form. We start with the classical Hamiltonian:
\begin{equation}
    H = \frac{a(t)}{2}\bigl(p^2 + x^2\bigr) + \frac{b(t)}{2}\bigl(xp + px\bigr).
    \label{eq:GenHamiltonian}
\end{equation}
The dynamics are conveniently described using the coordinate vector $\mathbf{z} = [x, p]^T$, allowing the Hamiltonian to take the matrix form:
\begin{equation}
    H = \tfrac{1}{2}\,\mathbf{z}^T \cdot \mathbf{M}(t)\cdot \mathbf{z}, \qquad
    \mathbf{M}(t) = 
    \begin{bmatrix} 
    a(t) & b(t) \\
    b(t) & a(t) 
    \end{bmatrix}.
\end{equation}
In this notation, Hamilton's equations read as
\begin{equation}
    \dot{\mathbf{z}} = \mathbf{J} \cdot \mathbf{M} \cdot \mathbf{z}, \qquad \mathbf{J} = \begin{bmatrix}0 & 1 \\ -1 & 0 \end{bmatrix},
\end{equation}
where $\mathbf{J}$ is the standard symplectic unit matrix. To factorize the time dependence, we introduce a linear canonical transformation $\mathbf{z}_{\mathbf{S}} = \mathbf{S}(t) \cdot \mathbf{z}$, which must preserve the symplectic structure ($\mathbf{S} \cdot \mathbf{J} \cdot \mathbf{S}^T = \mathbf{J}$). By choosing a lower triangular form for $\mathbf{S}(t)$ to decouple the coordinate scaling from the momentum displacement,
\begin{equation}
    \mathbf{S}(t) = \begin{bmatrix} A(t) & 0 \\ C(t) & D(t) \end{bmatrix}, \qquad
    \mathbf{z}_{\mathbf{S}} = \begin{bmatrix} x_{\mathbf{S}} \\ p_{\mathbf{S}} \end{bmatrix},
\end{equation}
we demand that the transformed Hamiltonian matrix $\mathbf{M}_{\mathbf{S}}$ becomes isotropic and stationary in a scaled time frame~\cite{Leach_1977,Leach_1978}:
\begin{equation}
    \mathbf{M}_{\mathbf{S}} = \frac{\omega_0}{\rho^2}\begin{bmatrix}1 & 0 \\ 0 & 1 \end{bmatrix}.
\end{equation}
Here, $\rho(t)$ is an auxiliary scaling function and $\omega_0 = \omega(t_0)$. The evolution of the transformation matrix is governed by $\dot{\mathbf{S}} = \mathbf{J} \cdot \mathbf{M}_{\mathbf{S}} \cdot \mathbf{S} - \mathbf{S} \cdot \mathbf{J} \cdot \mathbf{M}$. This matrix relation expands into the following set of coupled equations:
\begin{subequations} \label{eq:SystemODES_new}
    \begin{align}
    \dot{A} &= \frac{\omega_0}{\rho^2}C - b A, \\  
    aA &= \frac{\omega_0}{\rho^2}D, \\
    \dot{D} &= bD - aC, \\
    \dot{C} &= aD - bC - \frac{\omega_0}{\rho^2}A.
    \end{align}
\end{subequations}
Solving for $A, C,$ and $D$ in terms of $\rho(t)$ yields:
\begin{equation}\label{eq:xpClassical}
    x_{\mathbf{S}} = A x, \quad p_{\mathbf{S}} = C x + D p,
\end{equation}
with
\begin{equation}
\begin{split}
    A &= \sqrt{\frac{\omega_0}{a}}\frac{1}{\rho}, \\C &= \frac{1}{\sqrt{a \omega_0}}\left(b\rho - \frac{\dot{a}\rho}{2a} - \dot{\rho}\right),\\
    D &= \sqrt{\frac{a}{\omega_0}}\rho.
\end{split}
\end{equation}
The consistency of this mapping requires that $\rho(t)$ satisfies the Ermakov–Pinney equation:
\begin{equation}\label{eq:Ermakov}
    \ddot{\rho} + \omega^2(t)\rho = \frac{\omega_0^2}{\rho^3},
\end{equation}
with the time–dependent frequency defined as $\omega^2(t) = a^2 - b^2 - \dot{b} + \dot{a}b/a + \ddot{a}/(2a) - 3\dot{a}^2/(4a^2)$. 

In the transformed frame, the energy $H_{\mathbf{S}}$ assumes the isotropic form $\frac{\omega_0}{2\rho^2}(p_{\mathbf{S}}^2 + x_{\mathbf{S}}^2)$, from which we identify the action $P = \frac{1}{2}(p_{\mathbf{S}}^2 + x_{\mathbf{S}}^2)$ as an exact constant of motion. Returning to the original coordinates, we construct the Lewis–Ermakov invariant:
\begin{equation}\label{eq:I_HO}
\begin{split}
    I &= \frac{1}{2}\bigg\{\left[\frac{1}{\sqrt{a \omega_0}}\left( b\rho - \frac{\dot{a}\rho}{2a} - \dot{\rho}\right)x + \sqrt{\frac{a}{\omega_0}}\rho\,p\right]^2 \\&+ \frac{\omega_0}{a}\left(\frac{x}{\rho}\right)^2 \bigg\}.
\end{split}
\end{equation}
This classical framework provides the rigorous foundation for the operator formalism presented in the next section.

\section{Quantum Lewis–Ermakov invariant for the harmonic oscillator}\label{A_B}
The classical canonical transformation $\mathbf{z}_{\mathbf{S}} = \mathbf{S}\cdot\mathbf{z}$ derived in Appendix~\ref{A_A} is promoted to the quantum regime via time–dependent unitary operators. The objective is to map $\hat{H}_{\texttt{THO}}$ in Eq.~\eqref{H_THO} into an isotropic, autonomous form $\hat{H}_{\mathbf{S}}$ that possesses a manifest constant of motion. To faithfully replicate the classical scaling matrix $\mathbf{S}(t)$, we implement two successive transformations: first, a momentum displacement operator $\hat{G}(t)$ (which corresponds to the phase-space shear $C/D$), and second, a time–dependent scaling operator $\hat{T}(t)$ (which generates the spatial compression $A=1/D$):
\begin{subequations}
    \begin{align}
        \hat{G}(t) &= \exp\left[\frac{i}{2a}\left( \frac{\dot{a}}{2a} + \frac{\dot{\rho}}{\rho} - b \right)\hat{x}^2\right],\\
        \hat{T}(t) &= \exp\left[-\frac{i}{2}\ln\left(\sqrt{\frac{a}{\omega_0}}\rho\right)\bigl(\hat{x}\hat{p} + \hat{p}\hat{x}\bigr)\right].
    \end{align}
\end{subequations}
These operators transform the coordinates according to $\hat{G}\hat{T}\hat{x}\hat{T}^\dagger\hat{G}^\dagger = A\hat{x}$ and $\hat{G}\hat{T}\hat{p}\hat{T}^\dagger\hat{G}^\dagger = C\hat{x} + D\hat{p}$, where $A, C,$ and $D$ are exactly the classical parameters found in Eq.~\eqref{eq:xpClassical}. Under the composite unitary mapping $\hat{R} = \hat{G}\,\hat{T}$, the Hamiltonian transforms as $\hat{H}_{\mathbf{S}} = \hat{R}^\dagger\hat{H}\hat{R}-i\hat{R}^\dagger\partial\hat{R}/\partial t$~\cite{Huerta_2025}, yielding:
\begin{equation}\label{H_S_new}
\begin{split}
    \hat{H}_\mathbf{S} &= \frac{1}{2}\left[\frac{\omega_0}{\rho^2}\hat{p}^2 + \left(\rho\ddot{\rho} + \omega^2(t)\rho^2\right)\frac{\hat{x}^2}{\omega_0\rho^2}\right],\\
     &= \frac{\omega_0}{2\rho^2}\bigl(\hat{x}^2 + \hat{p}^2\bigr),
\end{split}
\end{equation}
where the Ermakov–Pinney equation $\ddot{\rho} + \omega^2(t)\rho = \omega_0^2/\rho^3$ ensures the isotropic form. Explicitly, the effective frequency $\omega^2(t)$ is given by:
\begin{equation}
    \omega^2(t) = a^2 - b^2 -\dot{b} + \frac{\dot{a}b}{a} + \frac{\ddot{a}}{2a} - \frac{3\dot{a}^2}{4a^2}.
\end{equation}
The Lewis–Ermakov invariant $\hat{I}_{\texttt{THO}}$ is then obtained by inversely transforming the stationary Hamiltonian back to the laboratory frame:
\begin{equation}\label{eq:I_HO_final}
\begin{split}
    \hat{I}_{\texttt{THO}}(t) &= \hat{R}(t)\left[\frac{\omega_0}{2}\left(p^2+\hat{x}^2\right)\right]\hat{R}^\dagger(t)
\end{split}
\end{equation}
where the resulting expression exactly matches the classical invariant of Eq.~\eqref{eq:I_HO} upon applying the standard canonical quantization rules $x \rightarrow \hat{x}$ and $p \rightarrow \hat{p}$. This operator is an exact dynamical invariant for the Hamiltonian $\hat{H}_{\texttt{THO}}(t)$ defined in Eq.~\eqref{H_THO}, meaning it satisfies the quantum invariant condition $d\hat{I}_{\texttt{THO}}/dt = \partial\hat{I}_{\texttt{THO}}/\partial t - i[\hat{I}_{\texttt{THO}}(t), \hat{H}_{\texttt{THO}}(t)] = 0$. This confirms that the quantum dynamics are governed by the same universal scaling logic derived classically.

\section{Lewis–Ermakov invariant for two–level systems}\label{A_C}
This appendix extends the Lewis–Ermakov formalism to time–dependent two–level systems by leveraging the algebraic isomorphism between the $su(1,1)$ and $su(2)$ Lie algebras, as introduced in Section~\ref{seccion_1}. The operators $\{\hat{x}^2, \hat{p}^2, \hat{x}\hat{p} + \hat{p}\hat{x}\}$ appearing in the quadratic Hamiltonian close under commutation to form the non-compact $su(1,1)$ Lie algebra, whereas the Pauli operators $\{\hat{\sigma}_-, \hat{\sigma}_+, \hat{\sigma}_z\}$ generate the compact $su(2)$ algebra. Although these algebras generate representations of different Lie groups, they share a complexification to $sl(2,\mathbb{C})$ and possess locally isomorphic algebraic structures, differing essentially only in their topological compactness. 

The construction of the Lewis–Ermakov invariant for two–level systems follows the same systematic procedure developed for the harmonic oscillator: temporal factorization via the transformation $\hat{G}(t)$, followed by a scaling to an isotropic form via the transformation $\hat{T}(t)$. In the $su(2)$ framework, the operator $\hat{\sigma}_- + \hat{\sigma}_+$ plays the role of $\hat{x}^2 + \hat{p}^2$, and $\hat{\sigma}_z$ acts as the analog of $\hat{x}\hat{p} + \hat{p}\hat{x}$.

We take the two–level Hamiltonian of Eq.~\eqref{H_TLS}, where $a(t)$ is the coupling strength and $b(t)$ is the detuning. The Pauli operators $\hat{\sigma}_\pm$ satisfy the $su(2)$ commutation relations $[\hat{\sigma}_z, \hat{\sigma}_\pm] = \pm 2\hat{\sigma}_\pm$, $[\hat{\sigma}_+, \hat{\sigma}_-] = \hat{\sigma}_z$, and the nilpotent condition $\hat{\sigma}_\pm^2 = 0$. 

Following the structure established for the quantum harmonic oscillator in Appendix~\ref{A_B}, we construct the invariant through two successive time–dependent transformations. The first transformation $\hat{G}(t)$ eliminates the detuning term proportional to $\hat{\sigma}_z$, and the second transformation $\hat{T}(t)$ performs a time–dependent scaling:
\begin{subequations}
    \begin{align}
        \hat{G}(t) &= \exp\left[\frac{i}{a}\left(\frac{\dot{\rho}}{\rho} + \frac{\dot{a}}{2a} +ib\right)\hat{\sigma}_-\right], \\
        \hat{T}(t) &= \exp\left[\ln\left(\sqrt{\frac{a}{\theta_0}}\rho\right)\hat{\sigma}_z\right],
    \end{align}
\end{subequations}
where $\rho(t)$ is again an auxiliary scaling function to be determined and $\theta_0=\theta(t_0)$. A crucial distinction arises here with respect to the bosonic case: due to the non–Hermitian nature of the lowering operator $\hat{\sigma}_-$ and its nilpotent character $\hat{\sigma}_-^2 =0$, the transformation $\hat{G}(t)$ is not unitary. While $\hat{T}(t)$ remains unitary since it is generated by the Hermitian operator $\hat{\sigma}_z$, the combined transformation $\hat{S}_{\texttt{TLS}}(t)=\hat{G}(t)\hat{T}(t)$ is generally non-unitary. Consequently, we must work with the inverse transformations $\hat{G}^{-1}$ and $\hat{T}^{-1}$ instead of their Hermitian conjugates. 

These transformations act on the Pauli operators according to:
\begin{subequations}
    \begin{align}
        \hat{G}\,\hat{T}\,\hat{\sigma}_-\,\hat{T}^{-1}\hat{G}^{-1}\ &= \frac{\theta_0}{a}\frac{1}{\rho^2}\hat{\sigma}_-, \\
        \hat{G}\,\hat{T}\,\hat{\sigma}_+\,\hat{T}^{-1}\hat{G}^{-1}\ &= \frac{a}{\theta_0}\rho^2\hat{\sigma}_+ - i\frac{\rho^2}{\theta_0}\left(\frac{\dot{\rho}}{\rho} + \frac{\dot{a}}{2a} +ib\right) \hat{\sigma}_z\notag \\
        &+ \frac{\rho^2}{\theta_0a}\left(\frac{\dot{\rho}}{\rho} + \frac{\dot{a}}{2a} +ib\right)^2 \hat{\sigma}_-, \\
        \hat{G}\,\hat{T}\,\hat{\sigma}_z\,\hat{T}^{-1}\hat{G}^{-1}\ &= \hat{\sigma}_z + \frac{2i}{a}\left(\frac{\dot{\rho}}{\rho} + \frac{\dot{a}}{2a} +ib\right)\hat{\sigma}_-.
    \end{align}
\end{subequations}
Under $\hat{S}_{\texttt{TLS}}(t) = \hat{G}(t)\,\hat{T}(t)$, the Hamiltonian maps to Eq.~\eqref{H_S}, structurally analogous to $\hat{H}_\mathbf{S} = \omega_0(\hat{x}^2 + \hat{p}^2)/(2\rho^2)$ for the oscillator in Appendix~\ref{A_B}. Consistency requires $\rho(t)$ to satisfy Eqs.~\eqref{Ermakov_TLS} and~\eqref{theta}. The emergence of imaginary terms in $\theta^2(t)$ is a direct consequence of the non-unitary nature of the transformation $\hat{G}(t)$. These imaginary contributions distinguish the two–level Ermakov–Pinney equation from its bosonic counterpart in Eq.~\eqref{eq:Ermakov}, and are essential to maintain consistency with the $su(2)$ algebraic structure. Comparing with the harmonic oscillator case, we observe that the real part of $\theta^2(t)$ changes sign in the $b^2$ term (from $a^2 - b^2$ for the oscillator to $a^2 + b^2$ for the two–level system), directly reflecting the metric distinction between the non-compact $su(1,1)$ and compact $su(2)$ Lie algebras.

Following the same inverse transformation procedure established for the quantum harmonic oscillator, the Lewis–Ermakov invariant for the two–level system is constructed as:
\begin{equation}
\begin{split}
\hat{I}_{\texttt{TLS}} &= \,\hat{S}_{\texttt{TLS}}(t)\,(\hat{\sigma}_+ + \hat{\sigma}_-)\,\hat{S}_{\texttt{TLS}}^{-1}(t),\\
    &=\,\frac{a\rho^2}{\theta_0}\hat{\sigma}_+ + \left[\frac{\theta_0}{a\rho^2} + \frac{\rho^2}{a\theta_0}\left(\frac{\dot{\rho}}{\rho} + \frac{\dot{a}}{2a} +ib\right)^2\right]\hat{\sigma}_-\\
 &-i  \frac{\rho^2}{\theta_0}\left(\frac{\dot{\rho}}{\rho} + \frac{\dot{a}}{2a} +ib\right)\hat{\sigma}_z,
 \end{split}
 \end{equation}
This quantity satisfies $d\hat{I}_{\texttt{TLS}}/dt=0$. The parallel construction of invariants for both the harmonic oscillator (Appendix~\ref{A_B}) and two–level systems (this appendix) rigorously demonstrates the universality of the Lewis–Ermakov formalism. Despite their distinct algebraic signatures ($su(1,1)$ versus $su(2)$) and the necessity of navigating through a non-unitary transformation for the spin case, both systems follow an identical constructive logic. Ultimately, the unified framework presented in Section~\ref{seccion_1} is firmly founded on this algebraic correspondence, enabling the systematic transfer of exact non-adiabatic solutions and control protocols between continuous-variable and discrete-variable quantum systems. 

%

\end{document}